\documentclass[fleqn,10pt]{wlscirep}
\usepackage[utf8]{inputenc}
\usepackage[T1]{fontenc}
\usepackage{lineno}

\title{Possible favored Great Oxidation Event scenario on exoplanets around M-Stars with the example of TRAPPIST-1e}

\author[1,2,*]{Adam Yassin Jaziri}
\author[1,3]{Nathalie Carrasco}
\author[4]{Benjamin Charnay}
\affil[1]{LATMOS/IPSL, UVSQ Universit\'{e} Paris-Saclay, Sorbonne Universit\'{e}, CNRS, Guyancourt, 78280, France}
\affil[2]{Laboratoire d’astrophysique de Bordeaux, Univ. Bordeaux, CNRS, B18N, all\'{e}e Geoffroy Saint-Hilaire, 33615 Pessac, France}
\affil[3]{ENS Paris-Saclay, Universit\'{e} Paris-Saclay, Gif-sur-Yvette, 91190, France}
\affil[4]{LIRA, Observatoire de Paris, Université PSL, CNRS, Sorbonne Université, Université Paris Cité, 5 place Jules Janssen, 92195 Meudon, France}

\affil[*]{contact: yassin.jaziri@latmos.ipsl.fr}

\begin{abstract}

The Great Oxidation Event (GOE), which marked the transition from an anoxic to an oxygenated atmosphere, occurred 2.4 billion years ago on Earth, several hundreds of millions of years after the emergence of oxygenic photosynthesis. This long delay implies that specific conditions in terms of biomass productivity and burial were necessary to trigger the GOE. It could be a limiting factor for the development of oxygenated atmospheres on inhabited exoplanets. In this study, we explore the specificities of a terrestrial planet in the habitable zone of an M dwarf for a GOE. Using a 1D coupled photochemical-climate model, we simulate the atmospheric evolution of TRAPPIST-1e, an Earth-like exoplanet, exploring the effect of oxygen sources (biotic or abiotic). Our results show that the stellar energy distribution promotes O$_3$ production at lower O$_2$ concentrations compared to Earth, and the ozone layer on TRAPPIST-1e forms more efficiently. This lowers the threshold for atmospheric oxidation, suggesting that the GOE on TRAPPIST-1e would occur quickly after the rise of oxygenic photosynthesis, up to 1Gyrs earlier than on Earth, and would reach O$_2$ enabling oxygenic respiration and thus the development of animals. We may question whether this is a general behavior around several M-stars. Furthermore, we discuss how the overproduction of ozone could make O$_3$ detection possible using the James Webb Space Telescope, providing a potential method to observe oxygenation signatures on exoplanets in the near future. Previous studies predicted that for an Earth-like atmosphere O$_3$ would require over 150 transits for detection, but our results show that significantly fewer transits could be needed.

\end{abstract}
\begin{document}

\flushbottom
\maketitle

\thispagestyle{empty}

Around 2.4 Gyr ago, the Earth's anoxic atmosphere faced the Great Oxidation Event (GOE) \cite{Catling2020}. During this event, the amount of oxygen increased from less than 10$^{-5}$ present atmospheric level (PAL) to a maximum of 10$^{-1}$ PAL around 2.2 Gyr ago \cite{Lyons2014}. A key indirect indicator of atmospheric O$_2$ concentration prior to the GOE is sulfur isotope fractionation (MIF-S). In the anoxic atmosphere of the Archean, MIF-S is favored in sedimentary rocks \cite{Farquhar2000, Pavlov2002}. However, after the GOE, MIF-S disappears, indicating a shift in atmospheric composition. Other geological indicators also provide constraints on the evolution of oxygen during the GOE. For example, the disappearance of detrital pyrite and uraninite in sedimentary rocks \cite{Johnson2014}, chromium isotope anomalies in Proterozoic ironstones \cite{Planavsky2014}, and the appearance of terrestrial red beds in the Paleoproterozoic rock record \cite{Shawwa2024} all serve as markers that attempt to constrain atmospheric O$_2$ levels. In addition to the few geological records, biogeochemical and photochemical models of the GOE give a possible understanding of this event \cite{Zahnle2006,Claire2006,Goldblatt2006,Gebauer2017,Gregory2021,Jaziri2022}. They show an atmospheric instability moving from low to high atmospheric O$_2$ levels with a sharp transition and triggered around 10$^{-5}$ surface O$_2$ level. This phenomenon arises from the formation of the ozone layer at high O$_2$ levels, which reduces the required biological O$_2$ input from the surface to sustain high atmospheric O$_2$. Previous studies suggest that O$_2$ loss through methane oxidation, which is the main O$_2$ loss in the atmosphere and catalyzed by OH radicals, decreases as a result of the UV shielding effect of O$_3$. This shielding limits H$_2$O photolysis, which produces the OH radicals catalyzing the methane oxidation \cite{Kasting1980,Pavlov2002,Goldblatt2006,Ruiz2023}. Another study suggests that O$_2$ loss via methane oxidation becomes less efficient due to the establishment of the Chapman cycle, which recycles O$_3$ back into O$_2$ \cite{Zahnle2006}. Interestingly, while all models point to a clear link between O$_3$ formation and the decrease of O$_2$ loss through methane oxidation, the explicit development of this connection remains unclear. Furthermore, the dynamics of the GOE differs across models due to substantial uncertainties and a lack of constraints on the biosphere at that time. Indeed, several hypotheses have been proposed to explain the timing of the GOE: hydrogen escape \cite{Catling2001}, emergence of continents and subaerial volcanism \cite{Gaillard2011}, secular evolution of the mantle redox state \cite{Kadoya2020}, and its consequences on volcanism outgassing fluxes \cite{O'Neill2022}.

The GOE was the first step in Earth's transition toward an aerobic biosphere and the development of complex life reliant on oxygenic respiration. This raises the question of how, when, and where a similar event might occur on habitable exoplanets and if the detection of O$_3$ could serve as a biosignature. Being able to observe such event and atmosphere could also provide valuable insights into Earth's past climate. Today, we are able to detect and characterize temperate planets as small as mini-Neptunes, such as K2-18 b \cite{Madhusudhan2023} or TOI-270 d \cite{Benneke2024}, and Earth-like planets such as TRAPPIST-1 e \cite{Gillon2017}. Thanks to state-of-the-art instruments such as the James Webb Space Telescope (JWST) or the future Habitable Worlds Observatory (HWO), more of these worlds will be characterized. TRAPPIST-1~e is a representative case study for characterizing exo-Earths, as it is an Earth-sized temperate planet orbiting an M-dwarf star, the most common type of star in the Milky Way. These stars are different from the Sun, redder and more active in the ultraviolet (UV). Nevertheless, TRAPPIST-1 is more active in UV than most M-dwarf stars \cite{Wilson2021}. UV radiation is the source of photochemical processes and complex chemistry that could lead to the formation of prebiotic chemistry. This difference in stellar irradiation can lead to different climatic behavior and evolution. To observe a GOE-like event on exoplanets, we need to ask whether it is possible, how it evolves and whether it is observable. Using a fully coupled 1D photochemical-climate model \cite{Jaziri2022}, we analyze the atmospheric oxygenation on TRAPPIST-1 e, an Earth-sized temperate planets around an M-dwarf star, and compare it to the Earth's GOE. This approach focuses on changes in the atmospheric bistability and its consequences for a GOE-like event, rather than changes in the biosphere within this new environment. For this reason, an Earth-like biosphere is assumed in this study, even though variations in far and near UV radiation around M-dwarfs could impact oxygenic photosynthesis, and despite the fact that the geochemical state and evolution of Earth-like exoplanets would have their own effects on the timing of a GOE-like event. This follows the same approach as \cite{Gebauer2018} for an Earth-like planet around AD Leo, another M-dwarf star. Our study focuses on a different M-dwarf stellar energy distribution (SED), a limitation noted in \cite{Gebauer2018} for future work and it has recently been shown by \cite{Cooke2023} that variations in M-dwarf SEDs indeed affect the resulting ozone column density. Our numerical approach also differs from \cite{Gebauer2018}, who imposed a fixed timing for the GOE using their lower derived oxygen flux to conclude on a potentially earlier GOE-like event around AD Leo. In contrast, we calculate the oxygen fluxes for TRAPPIST-1e directly to model the revised timing and quantify the potential shift of a GOE-like event on TRAPPIST-1e.

\section*{A new insight into the explanation of the GOE}

\begin{figure}[ht]
\centering
\includegraphics[width=\linewidth, trim= 0cm 0cm 0cm 1cm, clip]{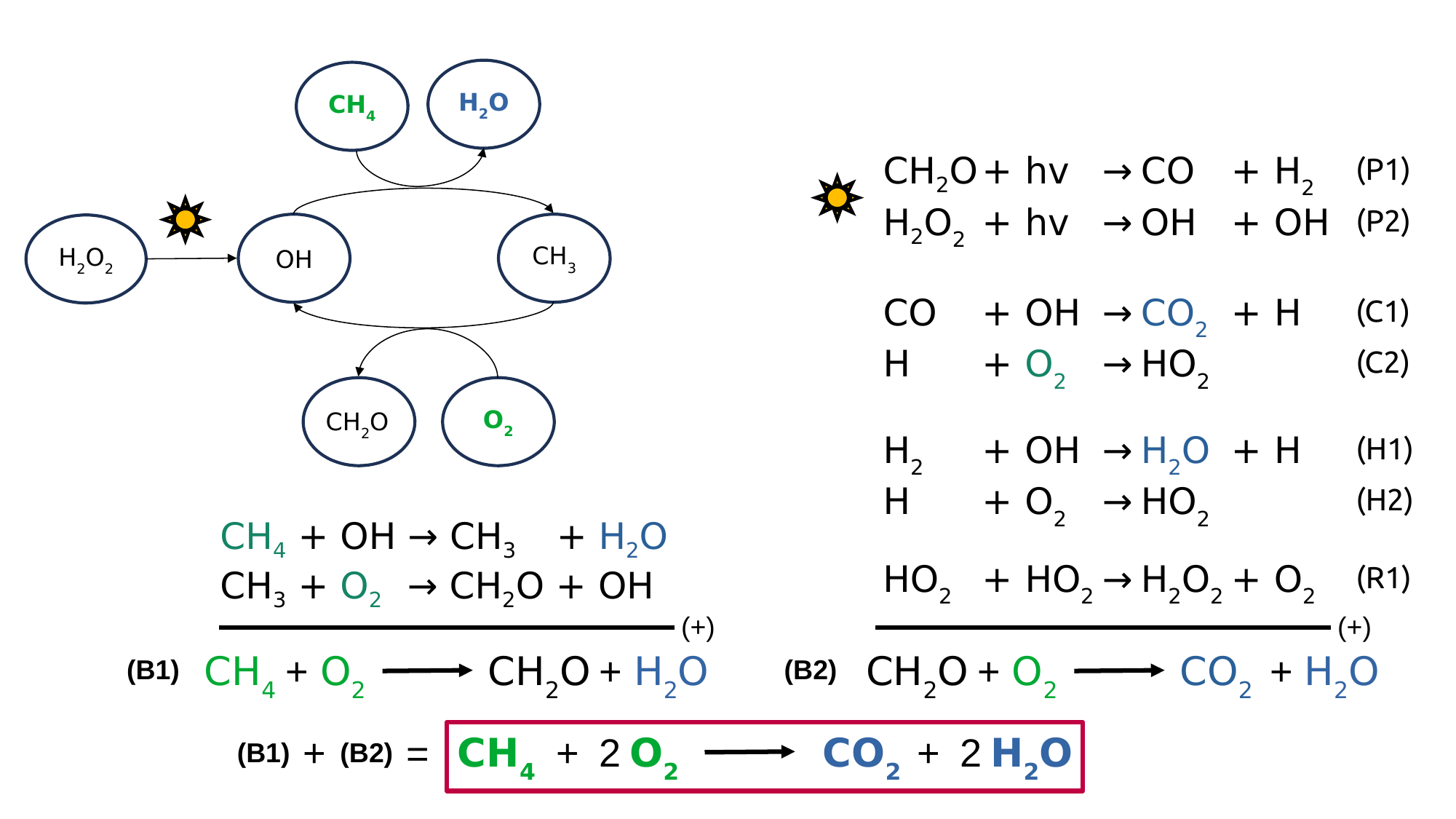}
\caption{Methane oxidation cycle and net reaction. The net reaction consumes methane to the benefice of CO$_2$ accumulation. However, this process actually involves a sequence of fast bimolecular reactions that interconvert short-lived radicals. These radicals are consumed as soon as produced, so that their concentrations remain stable but low in the atmosphere. They can be considered as catalysts of the net reaction. The catalytic cycle (B1) is illustrated on the figure: OH and CH$_3$ radicals are inter-converted, leading to the net consumption of CH$_4$ and O$_2$, and the production of CH$_2$O and H$_2$O.}
\label{fig:met_oxi}
\end{figure}

The loss of O$_2$ in the atmosphere on a photochemical timescale is primarily driven by methane oxidation \cite{Zahnle2006}. This net reaction is the result of a series of elementary reactions catalyzed by OH radicals generated through photochemical processes (see Fig. \ref{fig:met_oxi}). As O$_2$ levels rise, the efficiency of this reaction increases, leading to a higher flux of atmospheric O$_2$ loss. However, beyond a certain O$_2$ threshold, this flux begins to decrease due to a reduction in the production of OH radicals.

While H$_2$O photolysis has been widely recognized as the primary source of OH radicals \cite{Kasting1980,Pavlov2002,Goldblatt2006,Ruiz2023}, our model reveals that OH radicals are predominantly produced via the photolysis of H$_2$O$_2$ near the surface (reaction (P2)), as shown Fig. \ref{fig:oh_rates} (see appendix). All studies referring to this pathway are based on the findings of \cite{Kasting1980}, which show a clear correlation between the abundance of CH$_4$ and OH but do not provide a detailed pathway analysis justifying the origin of OH radicals from H$_2$O photolysis. H$_2$O$_2$, when photolyzed, acts as a catalyst for the reaction of methane oxidation. This finding aligns with previous detailed pathway analyses \cite{Gebauer2017}. The decline in O$_2$ loss flux and OH radical production (dominated by the photolysis of H$_2$O$_2$) is correlated with the formation of the ozone layer. The H$_2$O$_2$ photolysis occurs at wavelengths strongly absorbed by O$_3$, as shown in Fig. \ref{fig:stel_flux}. Consequently, the emergence of the ozone layer leads to O$_3$ shielding of H$_2$O$_2$ photolysis, reducing the production of OH radicals that drive methane oxidation. Additionally, the identification of formaldehyde photolysis (reaction (P1)) in the methane oxidation cycle further amplifies this O$_3$ shielding effect, because formaldehyde photodissociation also occurs at wavelengths absorbed by O$_3$, as shown in Fig. \ref{fig:stel_flux}.

As methane oxidation becomes less efficient, O$_2$ levels in the atmosphere increase, leading to further O$_3$ formation. This amplifies the shielding effect on H$_2$O$_2$ photolysis, reducing methane oxidation even further. This positive feedback loop triggers a runaway increase in O$_2$ levels until other processes intervene to stabilize atmospheric O$_2$. This sequence of events characterizes the GOE.

In summary, we identify the photochemical processes responsible for the GOE, highlighting the critical role of O$_3$ shielding H$_2$O$_2$ photolysis, limiting methane oxidation and driving the runaway increase in atmospheric O$_2$.

\section*{Enhancement of ozone in the atmosphere}

\begin{figure}[ht]
\centering
\includegraphics[width=\linewidth]{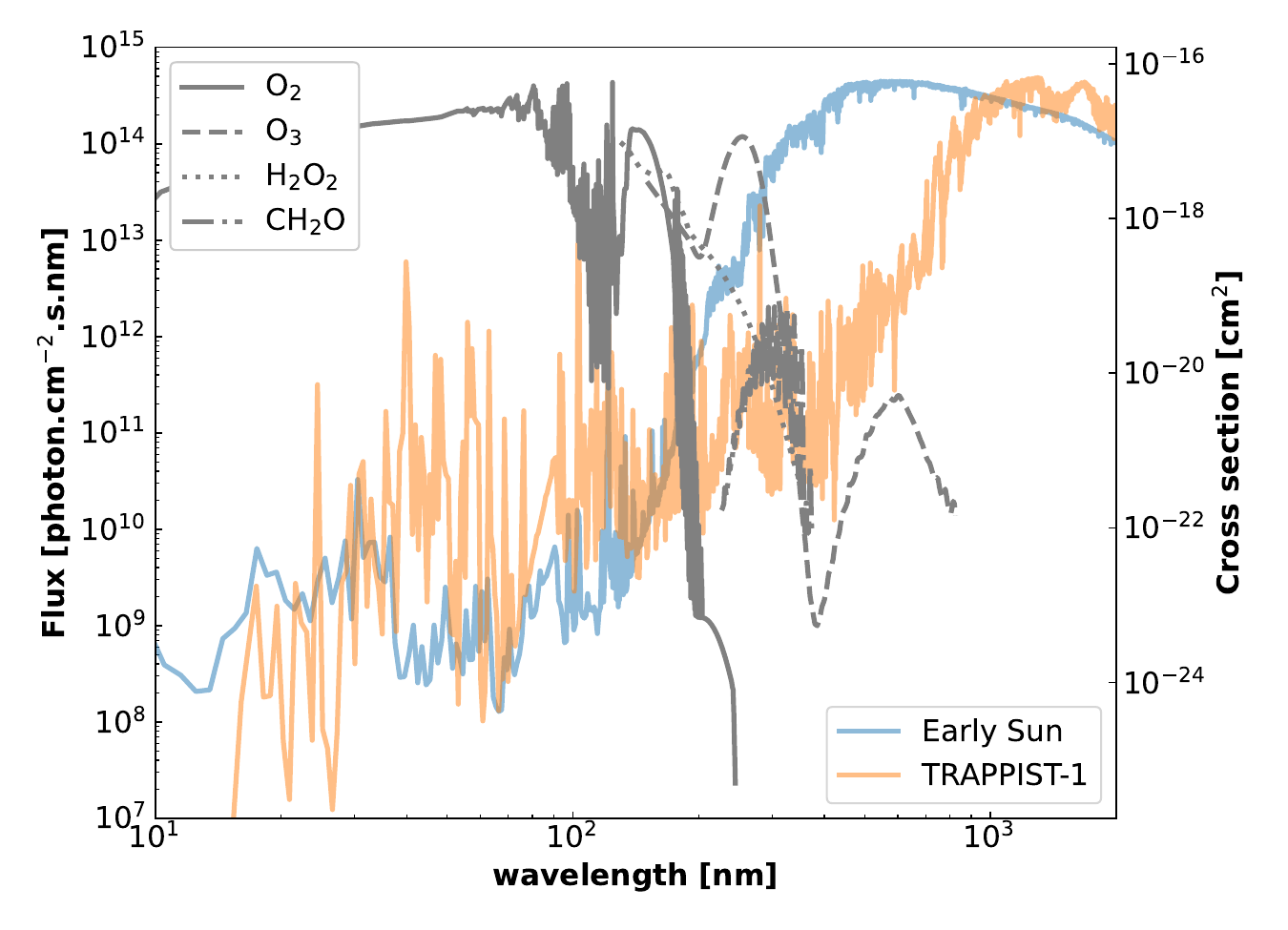}
\caption{Stellar irradiation received at the top of the atmosphere on Earth 2.7Ga ago (blue) \cite{Claire2012} and on TRAPPIST-1e (orange) \cite{Peacock2019}. Cross section per molecules of O$_2$, O$_3$ and H$_2$O$_2$ are represented in grey. Knowing O$_3$ less abundant than O$_2$, their photolysis shifting point is around 200nm, close to where the flux balance change between the Sun and TRAPPIST-1.}
\label{fig:stel_flux}
\end{figure}

To understand how the GOE would occur on TRAPPIST-1e, we used the Generic Planetary Climate Model (G-PCM). This is a Global Climate Model (GCM) that has recently been fully coupled in a generic way with photochemistry, and has been used to study the GOE of the early Earth \cite{Jaziri2022}. It can simulate in 1D and 3D the atmosphere of any type of planet \cite{Charnay2021,Turbet2021,Jaziri2022,Teinturier2024}. Here we simulate TRAPPIST-1e as a primitive Earth, around 2.4 Gyr ago, for a range of surface O$_2$ from 10$^{-8}$ to 10$^{-2}$. The parameters that have been changed, compare to the early Earth case \cite{Jaziri2022}, are the physical parameters \cite{Gillon2017}: radius, mass and rotation rate, and the incoming stellar flux (model 1A \cite{Peacock2019}) (Fig. \ref{fig:stel_flux}).

Previous work \cite{Cooke2023} explored different incident stellar fluxes under similar conditions and demonstrated that the relative UV flux, particularly the change in flux distribution between Earth and TRAPPIST-1e with a balance point around 200 nm (see Fig.~\ref{fig:stel_flux}), can favor a higher abundance of O$_3$ on TRAPPIST-1e compared to Earth. Within the framework of the Chapman cycle, this leads to an overall enhancement of odd oxygen on TRAPPIST-1e (see Fig.~\ref{fig:odd_oxygen} in appendix).

\begin{figure}[ht]
\centering
\includegraphics[width=\linewidth]{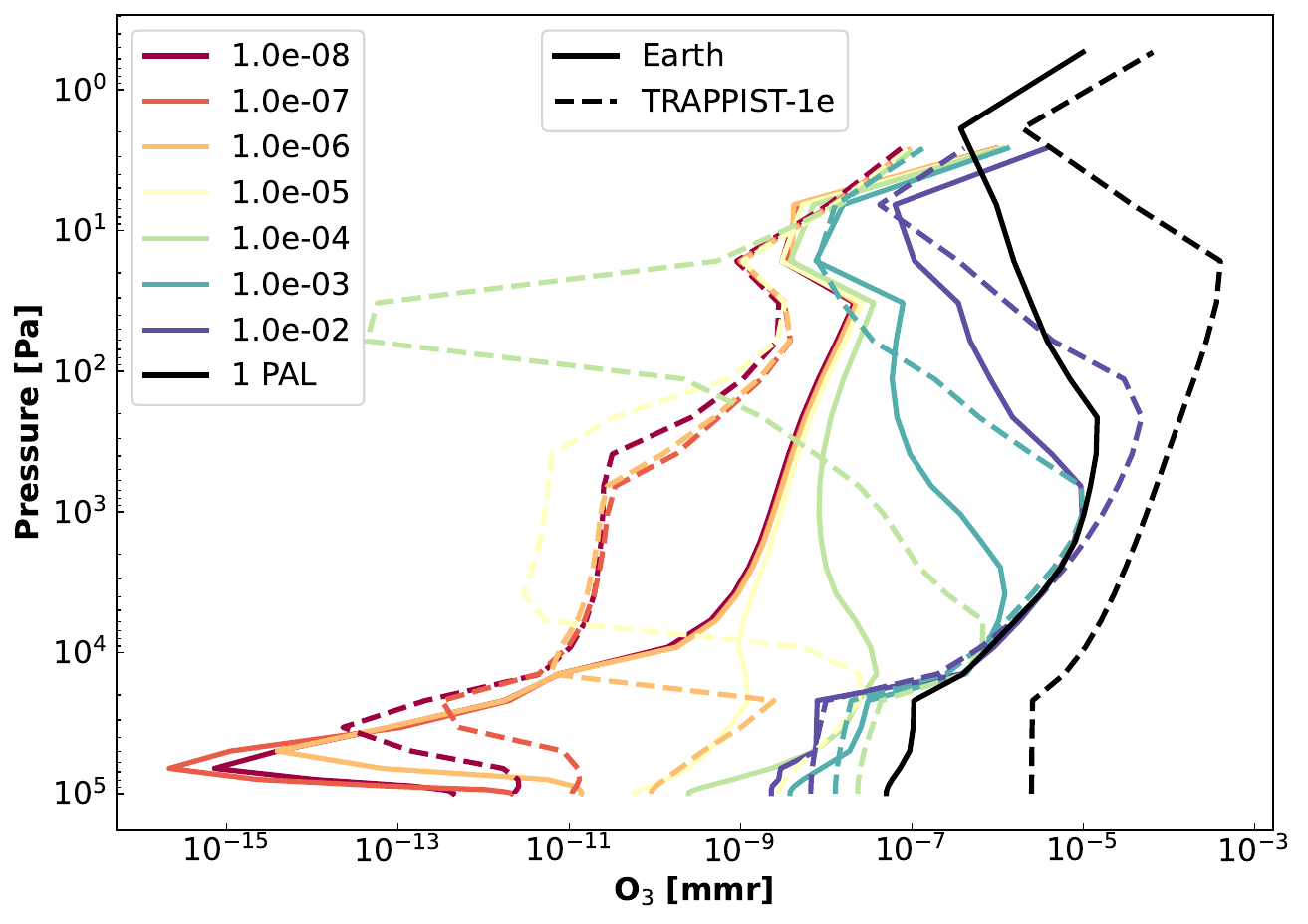}
\caption{Ozone profiles in mass mixing ratio (mmr) for different levels of surface O$_2$ from the 1D G-PCM. 1 PAL represent the present atmospheric level. Solid lines are early Earth simulations and Earth in black. Dashed lines are TRAPPIST-1e analogue simulations. Both planets differ mainly by their stellar irradiation. TRAPPIST-1e is favorable to the production of more ozone compare to Earth.}
\label{fig:o3_profiles}
\end{figure}

In this study, we adopt the stellar spectrum from \cite{Peacock2019} with the highest UV emission, consistent with models predicting enhanced UV emission in the early history of M-dwarf stars (remaining within roughly one order of magnitude evolution for TRAPPIST-1 \cite{Fleming2020}). As shown by \cite{Cooke2023}, this choice does not alter much the total ozone column density at low O$_2$ levels compare to present day, but it becomes more favorable for O$_3$ overproduction at 1 PAL O$_2$. Additionally, we neglect stellar activity in our simulations. The impact of stellar flares on exoplanets remains poorly understood but depending on their properties it can decrease or increase the abundance of O$_3$ \cite{Tilley2019,Ridgway2023}. We show in Fig.~\ref{fig:o3_profiles} that in this scenario O$_3$ abundance is enhanced on TRAPPIST-1e compared to Earth. O$_3$ profiles reveal that the ozone layer forms on TRAPPIST-1e at lower levels of O$_2$, and that ozone near the surface can reach concentrations potentially detrimental to the development of life \cite{Cooke2024}, challenging the assumption of an Earth-like biosphere.

\section*{Favored oxidation of the atmosphere around M-dwarf}

\begin{figure}[ht]
\centering
\includegraphics[width=\linewidth]{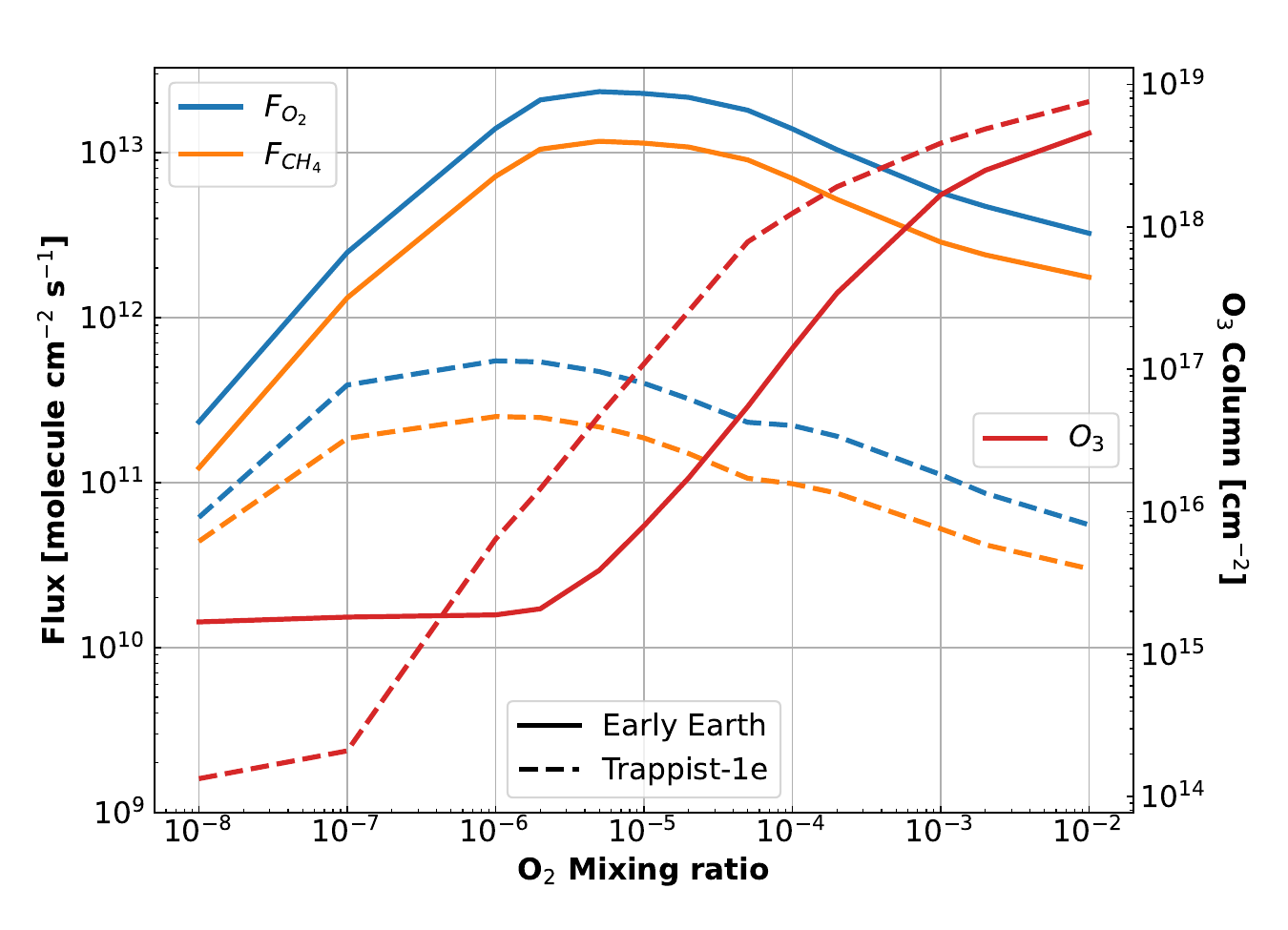}
\caption{Oxygen atmospheric loss ($F_{O_2}$), methane atmospheric loss ($F_{CH_4}$) and O$_3$ column density as a function of surface O$_{2}$. Surface abundances of [CH$_4$] = 10$^{-4}$ and [CO$_2$] = 10$^{-2}$. Results of early Earth model on Earth (solid) and on TRAPPIST-1e (dash).}
\label{fig:flux_surf}
\end{figure}

Atmospheric O$_2$ consumption is primarily driven by methane oxidation (Fig. \ref{fig:met_oxi}), which depletes O$_2$ and CH$_4$ in a 2:1 ratio. Fig. \ref{fig:flux_surf} illustrates the atmospheric loss fluxes of O$_2$ and CH$_4$, alongside the total O$_3$ column density, as modeled by the G-PCM, with varying surface abundances of O$_2$. The 2:1 ratio observed in these fluxes aligns with the stoichiometry of methane oxidation (Fig. \ref{fig:met_oxi}), the dominant mechanism for atmospheric O$_2$ consumption. As the surface abundance of O$_2$ increases, we observe a reduction in atmospheric loss fluxes, accompanied by a rise in total O$_3$ column density. This behavior is attributed to the UV shielding effect of O$_3$, as discussed in the previous section. Consequently, for a given rate of O$_2$ production by the biosphere, the system exhibits two stable states: one with low O$_2$ levels and another with high O$_2$ levels. In the past, Earth transitioned from a low O$_2$ state to a high O$_2$ state as exchange of oxidized and reduced materials gradually increased atmospheric O$_2$. Once the biosphere's O$_2$ net input surpassed the maximum atmospheric loss flux, a runaway in O$_2$ levels occurred, corresponding to the Great Oxidation Event (GOE).

\begin{figure}[ht]
\centering
\includegraphics[width=10cm, trim=0cm 1cm 0cm 0cm, clip]{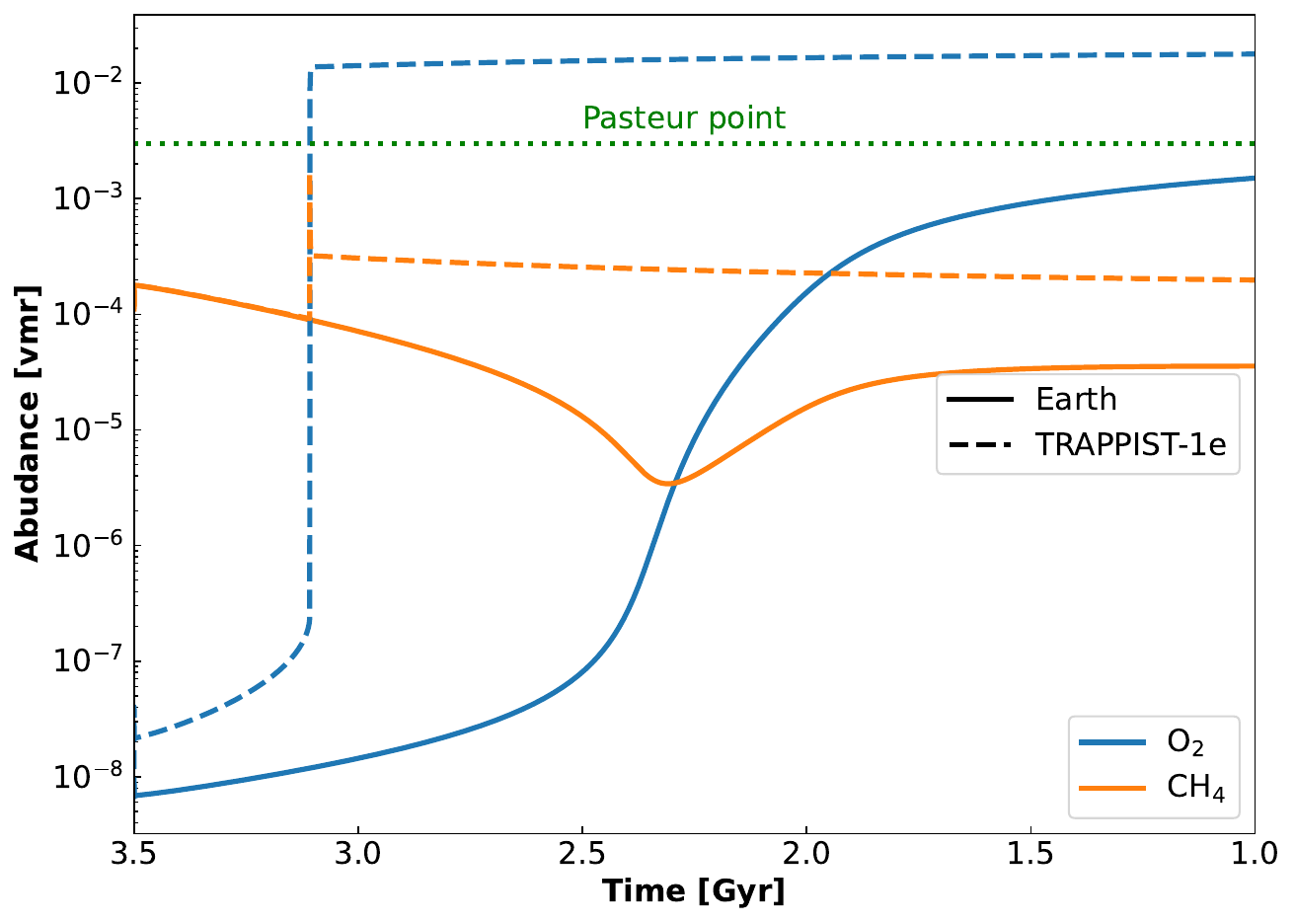}
\includegraphics[width=10cm]{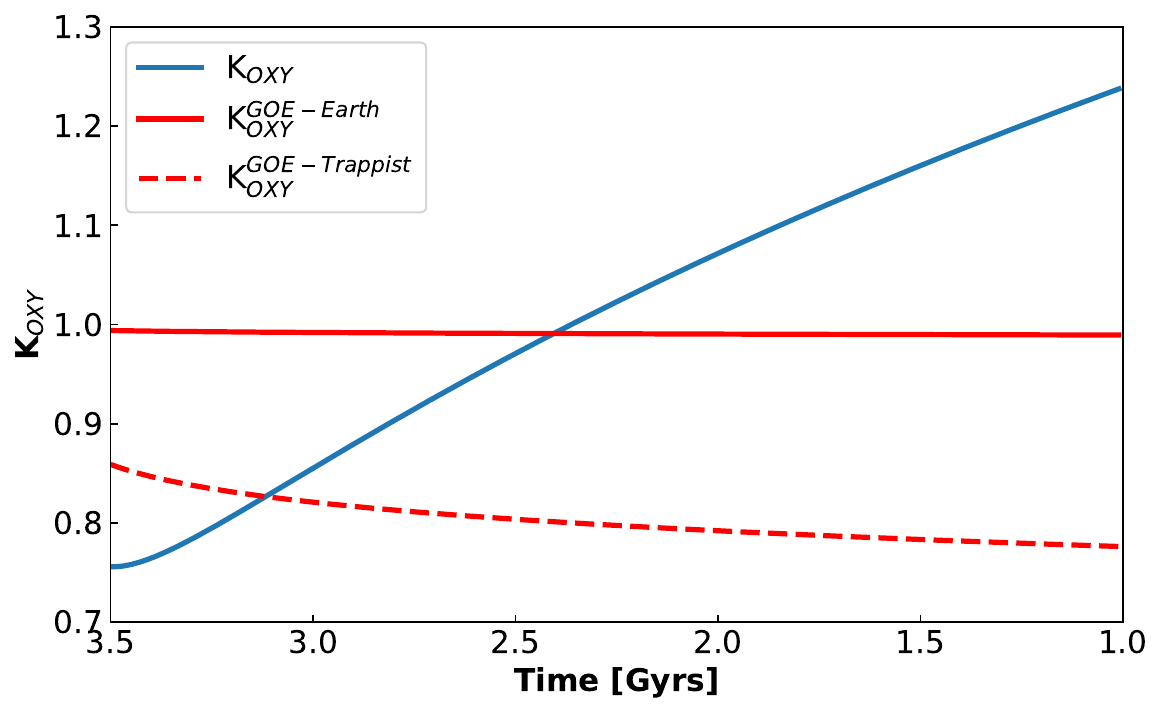}
\caption{Top: oxygen and methane abundance as a function of the time. Results of early Earth model on Earth \cite{Jaziri2022} (solid) and on TRAPPIST-1 e (dash). The Pasteur point (green pointed line) shows the oxygen level necessary to develop aerobic respiration. Bottom: evolution of the oxygenation parameter K$_{oxy}$=F$_{source}$/F$_{sink}$ at the surface (blue line). The critical value of K$_{oxy}$ corresponding to the triggering of the GOE is represented with a red line for the Earth (solid) and Trappist-1 e (dashed). K$_{oxy}$ and its critical value are expressed in the Methods section}
\label{fig:time_evolution}
\end{figure}

Fig. \ref{fig:flux_surf} compares the early Earth model \cite{Jaziri2022} with results for TRAPPIST-1 e model. Previous pathway analysis studies using similar modeling \cite{Gebauer2018} have shown that, due to higher UV radiation below 200 nm compared to Earth (see Figure \ref{fig:stel_flux}), the increased photolysis of H$_2$O shifts HOx from the troposphere to the upper atmosphere. HOx are key catalysts for methane oxidation (net loss of O$_2$) in the troposphere. As a result, global atmospheric loss fluxes are lower on TRAPPIST-1e. The most significant consequence of this reduced UV radiation above 200 nm is the favored formation of an ozone layer on TRAPPIST-1e. The enhanced O$_3$ levels on TRAPPIST-1e, relative to Earth, allow UV shielding to occur at lower O$_2$ concentrations, triggering the GOE earlier in the planet's evolution, assuming oxygenic photosynthesis emerges at the same time as it did on Earth. Using previous models of Earth's atmospheric evolution that account for all biosphere exchange of reduced and oxidized materials between the atmosphere-ocean system and the solid planet \cite{Jaziri2022}, the GOE on TRAPPIST-1e would occur, at most, approximately 1 Gyr earlier than on Earth (see Fig. \ref{fig:time_evolution}), coinciding in both cases with a total O$_3$ column density of $\sim$10$^{16}$ molecules.cm$^{-2}$ (see Fig. \ref{fig:flux_surf}). Fig. \ref{fig:time_evolution} also shows that, if the oxygenic photosynthesis is developed soon enough, the Pasteur point is reached earlier on TRAPPIST-1e, enabling the development of aerobic respiration shortly after the GOE, in contrast to Earth. 
The bottom panel of Fig. \ref{fig:time_evolution} shows the evolution of the oxygenation parameter K$_{oxy}$, defined as the surface flux ratio of oxygen source to oxygen sink \cite{Claire2006}. 
We derived an analytical formula of the critical value of K$_{oxy}$ above which the GOE is triggered (see Methods). For the Earth, the GOE occurs for K$_{oxy}$ $\sim$1 and at $\sim$2.4 Ga, so when surface oxygen source equals the surface oxygen sink. The photochemistry therefore had a small impact on the timing of the GOE. In contrast, the GOE for Trappist-1 e occurs for K$_{oxy}$ $\sim$0.83 and at $\sim$3.1 Ga, 700 Myrs before the GOE for Earth.
This shift in the timing of the GOE is driven by the spectral characteristics of M-dwarf stars that are similar to TRAPPIST-1, which possibly enhance ozone abundance, limit the atmospheric oxygen sink by methane oxidation, and thereby favor the oxidation of planetary atmospheres in their habitable zones. 

\section*{Exploring the potential for abiotic oxidation triggering}

Since it's easier to trigger oxidation on TRAPPIST-1e, the question arises as to whether lower abiotic oxygen fluxes could trigger atmospheric oxidation. The main known abiotic oxygen buildup is the water loss by hydrogen being subsequently lost through atmospheric escape due to the photolysis of H$_2$O \cite{Wordsworth2014,Luger2015,Lincowski2018}. Models of stellar evolution and atmospheric escape with different initial water conditions \cite{Krissansen2022,Gialluca2024} have predicted for TRAPPIST-1e an O$_2$ production rate of around $\sim$10$^{13}$ molecules.cm$^{-2}$.s$^{-1}$ during the first Gyr before decreasing significantly. While it doesn't reach a flux level of O$_2$ sufficient to trigger the GOE on Earth (see Fig. \ref{fig:flux_surf}), it is sufficient to trigger a GOE on TRAPPIST-1e. What's more, even though this level of flux only lasts for the first Gyr, it could be sufficient to trigger a GOE on TRAPPIST-1e, which can occurred earlier than on Earth (see Fig. \ref{fig:time_evolution}), if the oxygenic photosynthesis is developed as early.

Other abiotic sources of oxygen are known, such as the photochemical production of stable concentrations of O$_2$ from the photolysis of CO$_2$ \cite{Selsis2002,Domagal2014,Gao2015,Harman2015}. A recent study showed for TRAPPIST-1e \cite{Hu2020} that reaching 10\% CO$_2$ in a 1 bar N$_2$ atmosphere would trigger O$_2$ runaway. However, more recent work has raised the CO$_2$ threshold to a higher value \cite{Ranjan2023}. Another recent study has also identified a new pathway for the ionic production of O$_2$ from SO$_2$ \cite{Wallner2022}. This phenomenon remains to be quantified, for instance with atmospheric models and possible volcanic outgassing of SO$_2$.

\section*{Observing oxygenation signature}

\begin{figure}[ht]
\centering
\includegraphics[width=\linewidth]{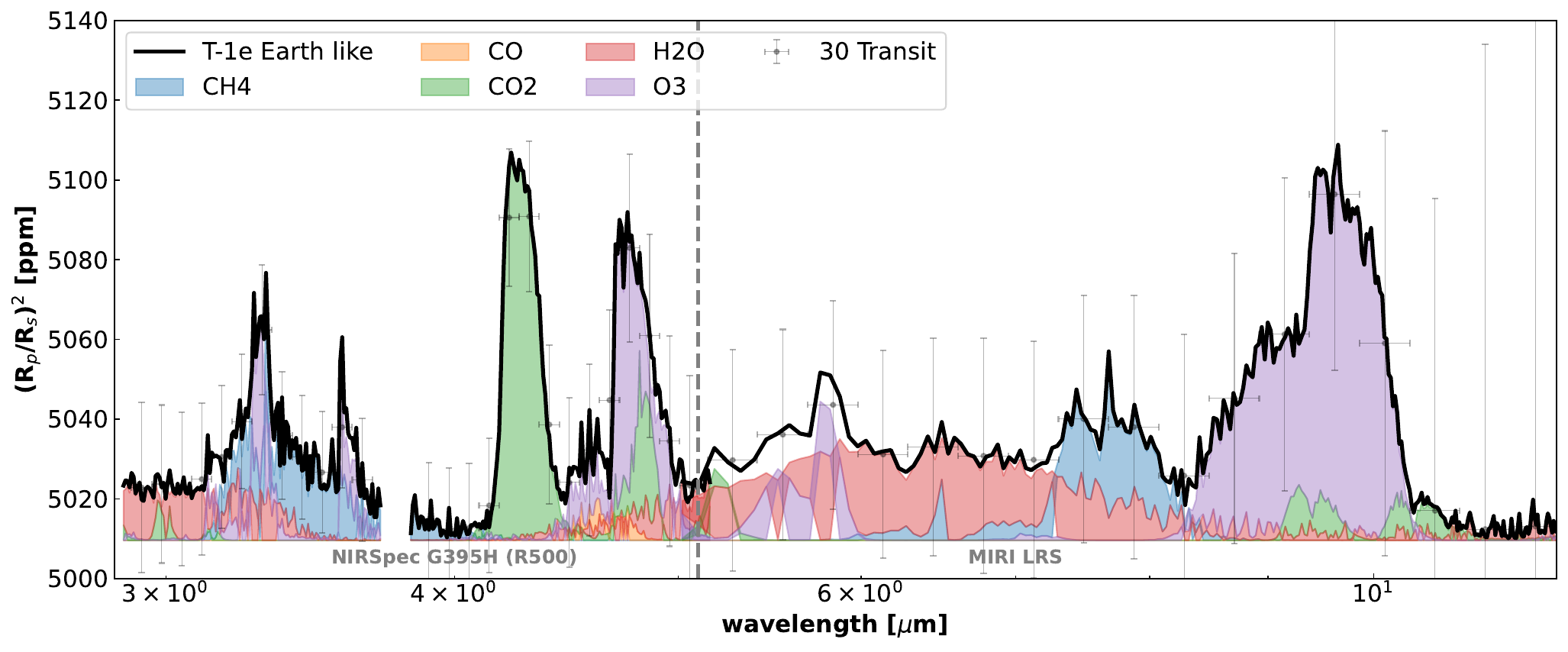}
\caption{TRAPPIST-1e transmission spectra for the Earth like simulation. Vertical dash line approximately separate JWST NIRSpec G395H range represented at a resolution of 500, and JWST MIRI LRS range. Simulated noise of JWST observations for 30 transits is represented with the error bars at a resolution of 50 and 20, respectively for NIRSpec and MIRI. Molecules contributions are represented with shaded colors. Main O$_3$ features are around 4.6 and 9.7 $\mu$m.}
\label{fig:spectra}
\end{figure}

Thanks to the unprecedented precision of the JWST, the detection of atmospheres on terrestrial planets orbiting small stars, such as the TRAPPIST-1 planets, is possible. Considering a case where TRAPPIST-1e has evolved in a similar way to Earth, we can verify the observability of O$_3$. As O$_3$ abundance is more pronounced on our TRAPPIST-1e scenario than on Earth (see Fig. \ref{fig:o3_profiles}), we can expect a stronger O$_3$ signature. Fig. \ref{fig:spectra} shows simulated transmission spectra, using the TauREx code \cite{Al-Refaie2021}, for an Earth-like atmosphere under TRAPPIST-1e conditions. The O$_3$ level is so high that, in addition to the main O$_3$ feature around 9.7 $\mu$m, which is the one we usually look to detect, we observe a strong O$_3$ feature around 4.6 $\mu$m. This falls within the range of the NIRSpec G395H JWST filter, which provides more precision than the MIRI LRS JWST filter, which would be used for the O$_3$ signature at 9.7 $\mu$m. Fig. \ref{fig:spectra} shows the error bars associated with 30 transit observations of these two JWST filters using the JWST noise simulator \cite{Batalha2017}. Although the O$_3$ feature at 9.7 $\mu$m is stronger than that at 4.6 $\mu$m, the instrument's uncertainty is higher. All this leads for the first time to the possibility of detecting O$_3$ using its signature at 4.6 $\mu$m. However, this detection requires moderate CO$_2$ levels. In our simulation, we assumed 1 PAL CO$_2$, corresponding to 4$\times$10$^{-4}$ vmr. Higher CO$_2$ levels would increase the number of transit observations needed to detect O$_3$ reliably.

The detection of minor gases, such as O$_3$, is a challenge. A previous study estimated 172 transits to detect O$_3$ on TRAPPIST-1e with its 9.7 $\mu$m signature and considering an atmosphere similar to Earth's \cite{Wunderlich2019}. Even with such a high number of transits, TRAPPIST-1e remains one of the best targets for observing O$_3$ on planets in the Habitable Zone around M-dwarf stars \cite{Wunderlich2019}. However, the overproduction of O$_3$ relative to Earth has not been taken into account. This not only improves the number of transits needed to detect O$_3$ with its feature at 9.7 $\mu$m, but also reveals an even lower number of transits with the new detectable O$_3$ feature at 4.6 $\mu$m. To verify the detection of O$_3$, we perform retrievals using TauREx \cite{Al-Refaie2021}, with and without O$_3$. Retrievals are performed on simulated TRAPPIST-1e transmission spectra, considering an Earth-like biosphere and JWST noise calculated for different numbers of transits. Simulated JWST spectra are calculated for the NIRSPec G395H and MIRI LRS instruments.

\begin{figure}[ht]
\centering
\includegraphics[width=\linewidth]{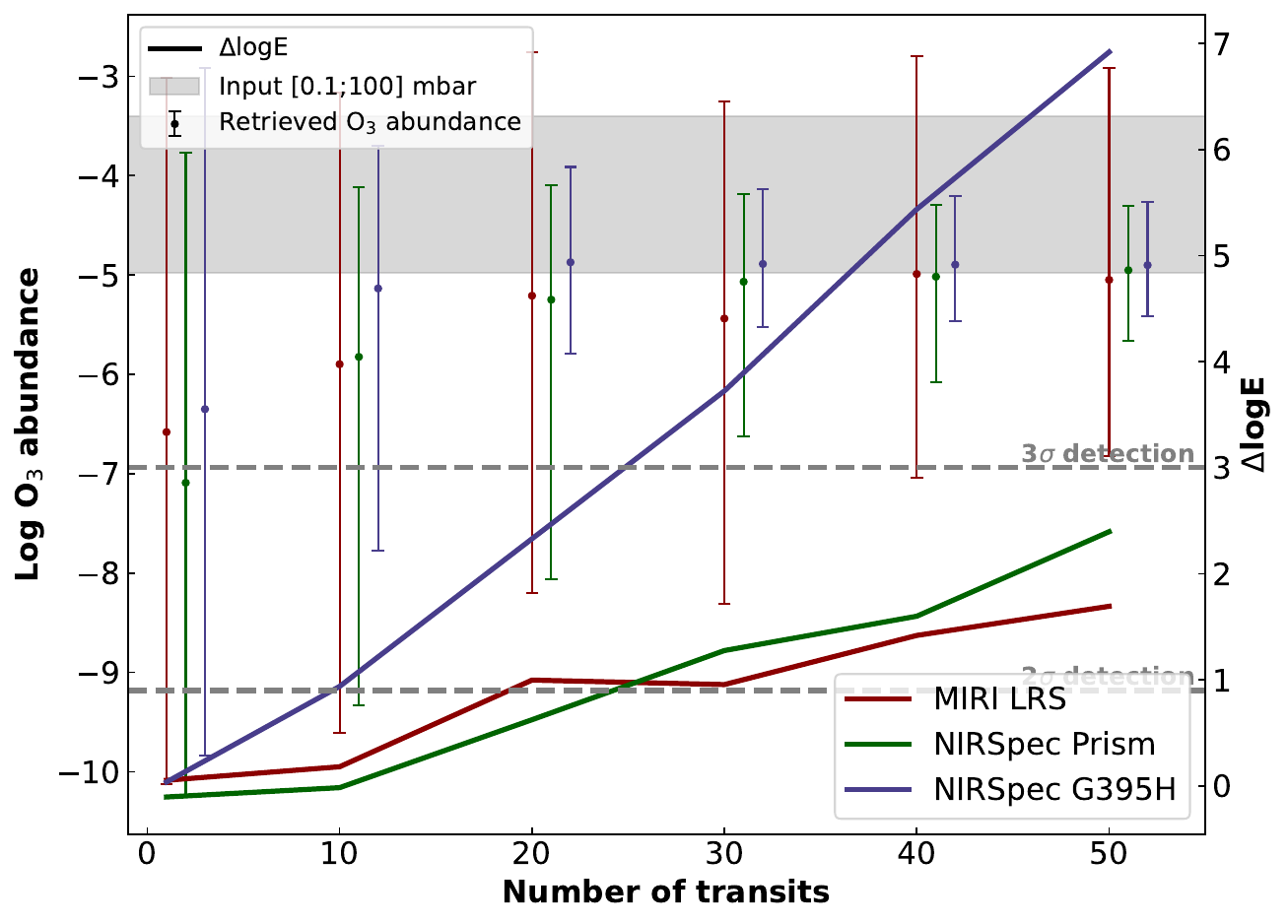}
\caption{Retrieval results for NIRSpec G395H (blue), NIRSpec Prism (green) and MIRI LRS (red) simulated observations as a function of the number of transit. O$_3$ retrieved abundances are plotted with error bars (NIRSpec Prism and G395H are shifted respectively by +1 and +2 transit for clarity). The modeled O$_3$ abundances between 0.1 and 100 mbar are represented with the shaded gray area. Its represents the atmosphere probed area. It is consistent with the retrieved values. Differential log evidence ($\Delta$logE) between retrievals considering or not O$_3$ are overplot. It represents the significance of the detection of O$_3$. 2$\sigma$ ($\Delta$logE=0.9) and 3$\sigma$ ($\Delta$logE=3.0) detection \cite{Benneke2013} values are plotted with horizontal dash lines. An O$_3$ 3$\sigma$ detection can be reach with 25 transits of NIRSpec G395H.}
\label{fig:retrieval}
\end{figure}

Fig. \ref{fig:retrieval} shows the retrieved O$_3$ abundance as a function of the number of transits. It is compared to the modeled values of O$_3$ abundance in the probed region of the atmosphere. As the number of transits increases, the uncertainty of the results decreases and the results converge toward the modeled value. The uncertainty in the O$_3$ abundance is 1 to 2 orders of magnitude lower with NIRSpec G395H than with MIRI LRS. To quantify detectability, we compare the logarithm of the evidence (logE) from retrievals with and without O$_3$. The difference, $\Delta$logE (Bayes Factor), is associated with the $\sigma$ detection levels for O$_3$ \cite{Benneke2013}. A good 3$\sigma$ detection is reached for $\Delta$logE = 3. As expected from the error bars, NIRSpec G395H provides a more reliable detection than MIRI LRS for the same number of transits. Fig. \ref{fig:retrieval} shows a linear trend of $\Delta$logE with the number of transits. We can extrapolate around 100 transits for a 3$\sigma$ detection with MIRI LRS, compared to the previous estimate of 172 transits \cite{Wunderlich2019}. This still represents an enormous number of transits required to detect O$_3$, which would also be the only molecule detectable by MIRI LRS. In contrast, NIRSpec G395H allows for a clear detection of CO$_2$ from 10 transits, and a reliable detection of O$_3$ and CH$_4$ from 20 transits. Fig. \ref{fig:retrieval} shows that a clear O$_3$ 3$\sigma$ detection with NIRSpec G395H can be achieved from 25 transits. These results drastically reduce the number of transits required to detect O$_3$ on TRAPPIST-1e, and for the first time provide an optimistic possibility for observing the oxygen-rich atmosphere of a terrestrial planet in the Habitable Zone.

The analysis was also conducted using NIRSpec Prism (spectra Fig. \ref{fig:spectra_bis} in appendix), which provides a broader wavelength range for characterizing the planet's atmosphere. However, its lower spectral resolution is insufficient compared to NIRSpec G395H. It is also important to note that these results do not account for stellar contamination, which is a significant factor for TRAPPIST-1. Considering this uncertainty, as well as the uncertainty in the adopted stellar spectrum, chosen to favor high O3 abundance at 1 PAL, the estimated number of required transits should be regarded as an optimistic lower limit. Furthermore, recent studies \cite{Welbanks2025,Fern2025} have emphasized the limitations of the Bayes factor criterion and its simplicity in determining detection significance, as well as potential molecular overlap. In our case, for example, the O$_3$ signature at 4.6 $\mu$m overlaps with that of CO$_2$, which would need to be disentangled observationally. Taking all these factors into account, a robust detection would likely require a 4–5$\sigma$ confidence level. A 4$\sigma$ detection would correspond to roughly 50 transits, still lower than previous estimates but notably more demanding than the optimistic case.

\section*{Conclusion}

The UV radiation of the TRAPPIST-1 spectrum used has been shown to influence atmospheric photochemistry in a way that favors the oxidation of planetary atmospheres within their habitable zones. This has also been found in a case study of AD Leo \cite{Gebauer2018}, raising the question of whether this is a general behavior for multiple M-stars. If so, while it took approximately 4 Gyrs on Earth to reach the Pasteur point, the threshold at which oxygenic respiration becomes energetically viable, this threshold could be reached much earlier following the rise of oxygenic photosynthesis on planets orbiting M stars. Consequently, environmental conditions favorable to the emergence of multicellular life could arise earlier on such planets. To confirm whether this is a general behavior, an extensive study using SEDs of M-stars from M0 to M9 of the MUSCLES (10.17909/T9DG6F) database should be conducted.

Although M-dwarf stars are often considered harsh environments due to tidal-locking, flares and their long pre-main sequence, these findings highlight their potential to support the emergence of complex life. However, it is important to remember that this is from an atmospheric photochemical perspective, and the possibility and efficiency of photosynthesis around M-dwarf stars still need to be better understood and characterized, along with the geochemical state and evolution of these environments.

\section*{Methods}

\subsection*{Photochemical-Global Climate Model}

The photochemical-climate model used to simulate the atmosphere of TRAPPIST-1e is the Generic Planetary Climate Model (G-PCM), a versatile 1D/3D model designed to efficiently handle a wide range of atmospheric conditions. The photochemical module was recently updated to be fully generic and more user-friendly \cite{Jaziri2022}, and has been tested on Earth, including studies of the Great Oxidation Event (GOE) of the early Earth \cite{Jaziri2022}. The model incorporates a detailed methane chemistry, which accounts for the various pathways of methane oxidation (\ref{fig:met_oxi}). Methane oxidation is the primary pathway for O$_2$ depletion in the early Earth's atmosphere, influencing its overall oxygen content. Understanding this process is crucial for modeling oxygenated planetary atmospheres.

We applied the same model and methodology used in the previous early Earth study \cite{Jaziri2022} to investigate the GOE on TRAPPIST-1e. The solar zenith angle is 60$^{\circ}$, the temperature is self-consistently calculated and the eddy diffusion coefficient is the same as previous analogue study \cite{Zahnle2006}. The same chemical network was used, without including the most recent updated parameterizations in the Schumann-Runge bands \cite{Ji2024}. These updates could be expected to change the quantitative results, such as the ozone abundance \cite{Ji2023}, but would not affect the qualitative outcomes. The surface O$_2$ levels has been varied between 10$^{-8}$ and 10$^{-2}$ while CH$_4$ was held constant at 10$^{-4}$. Given the consistent trends observed across different CH$_4$ levels, we chose a mid-range value for our simulations. Additionally, the planet's mass, radius and planet rotation were adjusted to match those of TRAPPIST-1e \cite{Agol2021}, as was the stellar irradiation \cite{Peacock2019}, as shown in Fig. \ref{fig:flux_surf}.

While a previous study demonstrated that both 1D and 3D simulations yield similar results for the photochemical oxidation flux on early Earth \cite{Jaziri2022}, we ran both 1D and 3D simulations for TRAPPIST-1e to assess the impact of its synchronous rotation. For clarity, we focus here on the results from the 1D simulations, which capture the global effects on the GOE that will be qualitatively the same with the results from the 3D simulations. The detailed impact of synchronous rotation on the O$_3$ abundance and O$_2$ surface flux distribution, as observed in the 3D simulations, will be explored in a future publication.

Finally, we simulated TRAPPIST-1e under Earth-like conditions to evaluate the potential observability of an oxygenated world. In this case, we reduced the levels of CO$_2$ to 4$\times10^{-4}$ and CH$_4$ to 1.8$\times10^{-6}$, set O$_2$ to 21\%, and replaced the detailed methane chemistry with NOx chemistry, in order to reflect the conditions on Earth.

\subsection*{Time evolution model}

The time evolution model of the abundances of O$_2$ and CH$_4$ is the same as the one developed in a previous work to study the early GOE of Earth \cite{Jaziri2022}. This model combines a work that parametrizes the time evolution equations of O$_2$ and CH$_4$ abundances \cite{Goldblatt2006} and a work that parametrizes the temporal evolution of redox flux at the surface of the planet \cite{Claire2006}. The equations are shown below, and full description can be found in Table \ref{tab:eqdynterms}.

\begin{equation}
\frac{d[CH_4]}{dt}=\frac{1}{2}\Omega_{O_{2}}N + \frac{1}{2}\Omega_{O_{2}}r - k_{\rm esc}[CH_4] - \frac{1}{2}\Psi_{O_{2}}[CH_4]^{0,7}% -  \frac{1}{2}\Omega_{O_{2}}(\beta(N+r)-wC)
\label{eq:dynM}
\end{equation}
\begin{equation}
\frac{d[O_2]}{dt}=\Omega_{O_{2}}N - (1-\Omega_{O_{2}})r - k_{\rm esc}[CH_4] - \Psi_{O_{2}}[CH_4]^{0,7}% - (1-\Omega_{O_{2}})(\beta(N+r)-wC)
\label{eq:dynO}
\end{equation}

The photochemical oxidation parametrization have been adapted for TRAPPIST-1e according to the atmospheric modeling results. Corrections factors, see Table \ref{tab:eqdynterms}, have been added to fit the new calculated O$_2$ flux, see Fig. \ref{fig:flux_surf}. Otherwise, all other parameters have been kept the same as the early Earth study to mimic an Earth-like biosphere and focus on the photochemical effects.

\begin{table}[ht]
\centering
\begin{tabular}{ccc}
\hline
\hline
Terms & Description & Values \\\hline
\multicolumn{3}{c}{\textbf{Atmospheric fluxes}}\\\hline
$\Psi_{O_{2}}[CH_{4}]^{0,7}$ & Photochemical oxidation (Earth) & $\Psi_{O_{2}} = 10^{0.0030\psi^{4}-0.1655\psi^{3}+3.2305\psi^{2}-25.8343\psi+71.5398}$ \\
 & & with $\psi = log([O_2])$ \\
 & Photochemical oxidation (TRAPPIST-1e) & $\Psi_{O_{2}} = 10^{0.0030\psi^{4}-0.1655\psi^{3}+3.2305\psi^{2}-25.8343\psi+70.5398}$ \\
 & & with $\psi = 1+log([O_2])$ \\
$k_{\rm esc}[CH_{4}]$ & Atmospheric escape & $k_{\rm esc} = 2.03\times10^{-5}$ $yr^{-1}$\\\hline
\multicolumn{3}{c}{\textbf{Surface fluxes}}\\\hline
$N$ & Oxygenic photosynthesis & $3.75\times10^{14}$ mol O$_2$ equiv.$yr^{-1}$\\
$\Omega_{O_{2}}$ & Fraction of O$_2$ produced reaching the atmosphere & $\Omega_{O_{2}} = (1-\gamma)(1-\delta)$\\
$\gamma$ & Fraction consumed & $\gamma = [$O$_2]/(d_{\gamma} + [$O$_2]) $\\
 & by heterotrophic respirers & $ d_{\gamma} = $ $1.36\times10^{19}$ mol \\
$\delta$ & Fraction consumed & $\delta = [$O$_2]/(d_{\delta} + [$O$_2]) $\\
 & by methanotrophs & $ d_{\delta} = $ $2.73\times10^{17}$ mol \\
\hline
$r$ & Net reducing surface flux & $F_{V}+F_{M}+F_{W}-F_{B}$ \\
 & Anoxygenic photosynthesis & mol O$_2$ equiv.$yr^{-1}$ \\
$F_{V}$ & Volcanic flux of reductants & $1.59\times10^{11}\left(\frac{3.586}{3.586-t}\right)^{0.17}$ \\
$F_{M}$ & Metamorphic outgassing of reductants & $6.12\times10^{11}\left(\frac{4.11}{4.11-t}\right)^{0.7}$ \\
$F_{B}$ & Burial & $1.06\times10^{12}\left(\frac{3.659}{3.659-t}\right)^{0.2}$ \\
$F_{W}$ & Oxidative weathering & $3.7\times10^{4}[O_2]^{0.4}$ \\
\hline
\hline
\end{tabular}
\caption{Equation dependencies and values \cite{Goldblatt2006}. Redox flux model \cite{Claire2006} with time \textit{t} [Gyrs] and amount of oxygen $[O_2]$ [mol]. \cite{Jaziri2022}}
\label{tab:eqdynterms}
\end{table}

\subsection*{Analytical expression of the timing of the GOE}

At steady-state and at low oxygen levels: $\frac{d[CH_4]}{dt}=\frac{d[O_2]}{dt}=0$, $\Omega_{O_{2}}\sim 1$ and $F_{M}\sim 0$. The equations of Table \ref{tab:eqdynterms} are equivalent to the equations from Claire et al. 2006\cite{Claire2006}, but with a different parametrization for photochemical oxidation. Assuming $N>>r$, with $r=F_{V}+F_{M}-F_{B}$, the steady-state abundances of $O_{2}$ and $CH_{4}$ are given by:
\begin{equation}
\frac{N}{2} - k_{\rm esc}[CH_4] -\frac{1}{2}\Psi_{O_{2}}[CH_4]^{0,7} = 0
\label{eq:eqM}
\end{equation}
\begin{equation}
N - k_{\rm esc}[CH_4] -\Psi_{O_{2}}[CH_4]^{0,7} + F_{B} -(F_{V}+F_{M})= 0
\label{eq:eqO}
\end{equation}
The oxidation parameter is defined as $K_{oxy}=\frac{F_{source}}{F_{sink}}\sim\frac{F_{B}}{F_{V}+F_{M}}$ \cite{Claire2006}. The abundances of CH$_4$ before the GOE can then be expressed as:
\begin{equation}
[CH_4]=\frac{F_{V}+F_{M}-F_{B}}{k_{esc}}=\frac{F_{B}}{k_{esc}}\left(\frac{1}{K_{oxy}}-1 \right)
\label{eq:eqMabundance}
\end{equation}
By subtracting equation (\ref{eq:eqM}) from equation (\ref{eq:eqO}) to eliminate $k_{\rm esc}[CH_4]$ and by substituting the expression of $[CH_4]$ from (\ref{eq:eqMabundance}), we get an expression of $K_{oxy}$:
\begin{equation}
K_{oxy}=\frac{1}{1+\frac{F_{B}}{k_{esc}}\left(\frac{N}{2\Psi_{O_{2}}} \right)^{1/0.7}}
\end{equation}
We define the critical value of $K_{oxy}$ at which the GOE is triggered as:
\begin{equation}
K_{oxy}^{GOE}=\frac{1}{1+\frac{F_{B}}{k_{esc}}\left(\frac{N}{2\Psi_{O_{2}}^{max}} \right)^{1/0.7}}
\end{equation}
where $\Psi_{O_{2}}^{max}$ is the maximal value of $\Psi_{O_{2}}$ and corresponds to the triggering of the oxygen instability and to strong increase of $O_2$ abundance until it is balanced by oxidative weathering. We note that $K_{oxy}^{GOE}<1$. This means that the atmosphere is always oxygenated for $K_{oxy}>1$, and that it could be oxygenated for $K_{oxy}<1$, especially for low values of $\Psi_{O_{2}}^{max}$ (i.e. for planets around M stars).
With our values from Table \ref{tab:eqdynterms}, $K_{oxy}^{GOE} =0.99$ at 2.4 Ga for Earth and $K_{oxy}^{GOE} \sim 0.83$ at 3.1 Ga for Trappist-1 e. This confirms the early GOE for Trappist-1 e relative to Earth, assuming the same surface redox fluxes.

\subsection*{Simulated JWST transit observations}
\subsubsection*{Forward model}

\texttt{TauREx} (Tau Retrieval for Exoplanets)\footnote{\url{https://github.com/ucl-exoplanets/TauREx3_public}} is an open-source Python tool designed for exoplanet atmospheric analysis. It operates with two primary Bayesian frameworks: Forward Model and Retrieval. Developed since many years \cite{waldmann2015a,waldmann2015b}, we used the latest version, \texttt{TauREx 3} \cite{Al-Refaie2021}. The Forward Model framework is capable of computing a 1D atmosphere or accepting one as input. In this study, we used the output from the G-PCM simulations as input. The tool then generates theoretical transmission or emission spectra at the desired spectral resolution, using molecular cross sections. Here, we used line-by-line cross sections computed by ExoMol \cite{Chubb2021}, and also accounted for Rayleigh scattering contributions.

For the specific case of TRAPPIST-1e with an Earth-like atmosphere, we calculated the theoretical transmission spectra. The temperature-pressure (T-P) profile and chemical abundances are taken from the G-PCM simulations, and system parameters are listed in Table \ref{tab:retrieval}. Spectra were generated at the spectral resolutions corresponding to JWST's NIRSpec G395H and MIRI LRS filters, extracted from the JWST noise simulatior Pandexo \cite{Batalha2017}, that ensure the consistency of our analysis. See results of the forward models Fig. \ref{fig:spectra}.

\subsubsection*{Noise simulator}

To simulate JWST observables (spectra Fig. \ref{fig:spectra}), we used the \texttt{PandExo} 1.5 package \cite{Batalha2017}, a noise simulator specifically designed for JWST exoplanet observations. We simulated 1 to 50 transits using NIRSpec G395H F290LP SUB2048 and MIRI LRS Slitless. A saturation limit of 80\% of the full well was applied and we considered a fraction of time out-of-transit to in-transit of 2. For each instrument, we selected the \textit{optimize} option for the number of groups per integration, which automatically determines the best settings to carry the observations. The system parameters used for producing the observables can be found in Table~\ref{tab:retrieval}. No additional random noise was included in the simulations.

\subsection*{Retrieval model}

We used the \texttt{TauREx 3} retrieval framework employing the nested sampling retrieval algorithm Multinest with its Python version \textit{PyMultiNest} \cite{feroz2009}. \texttt{TauREx 3} is capable of handling various retrieval models, including isothermal temperature profiles and multi-PT points, constant species profiles, 2-layers species profiles \cite{Changeat2019}, and chemical equilibrium models, which have been compared in previous studies \cite{Al-Refaie2022,Jaziri2024}. For this analysis, we focus on isothermal retrievals with constant species profiles, as these are computationally efficient and sufficient for qualitative species detection using transmission spectroscopy. The retrieval parameters and priors are summarized in Table \ref{tab:retrieval}.

\texttt{TauREx 3} is a full Bayesian retrieval framework that returns not only the best-fit transmission model spectrum but also the posterior distributions of all model parameters and the Bayesian evidence. To compare models, we use the Bayesian evidence \cite{Trotta2008, waldmann2015a} to compute the logarithmic Bayes factor:

\begin{equation}
  \Delta logE = \log{\frac{E_{model_A}}{E_{model_B}}} = \log{E_{model_A}} - \log{E_{model_B}}
  \label{eq:logbayes_factor}
\end{equation}

where $E_{model_A}$ and $E_{model_B}$ represent the evidences of two competing models. These Bayes factors are then translated into statistical significance \cite{Benneke2013}. A molecule is considered detected at a 3$\sigma$ level if $\Delta$logE $\geq$ 3.0.

We performed retrievals on each JWST simulated spectrum, computed by varying the number of transits. The retrievals were conducted both with and without O$_3$ as a retrieved species. The objective was to assess the detectability of O$_3$ as a function of the number of transits. By comparing the evidence from retrievals with and without O$_3$, we calculated the logarithmic Bayes factor to quantify the strength of the detection. See results Fig. \ref{fig:retrieval}.

\begin{table}[ht]
\centering
\begin{tabular}{c|c}
\hline
\hline
\textbf{Parameters}          & \textbf{Value}  \\
\hline
\multicolumn{2}{c}{\textbf{TRAPPIST-1}}        \\
\hline
T$_*$ [K]                    & 2566.0          \\
R$_*$ [R$_\odot$]            & 0.12            \\
Z$_*$                        & 0.04            \\
log \textit{g} [cm.s$^{-2}$] & 5.24            \\
Magnitude band J             & 11.354          \\
\hline
\multicolumn{2}{c}{\textbf{TRAPPIST-1e}}       \\
\hline
R$_p$ [R$_\oplus$]           & 0.920           \\
M$_p$ [M$_\oplus$]           & 0.692           \\
Transit duration [days]      & 0.03910417      \\
\hline
\multicolumn{2}{c}{\textbf{Retrievals priors}} \\
\hline
Radius [R$_{jup}$]           & 0.080 to 0.085  \\
T [K]                        & 50 to 350       \\
log$_{10}$(H$_2$O)           & -12 to -1       \\
log$_{10}$(CO)               & -12 to -1       \\
log$_{10}$(CH$_4$)           & -12 to -1       \\
log$_{10}$(CO$_2$)           & -12 to -1       \\
\hline
log$_{10}$(O$_3$)            & -12 to -1       \\
\hline
\hline
\end{tabular}
\caption{\label{tab:retrieval}System parameters \cite{Agol2021}, retrievals free parameters and priors.}
\end{table}

\bibliography{biblio}

\section*{Research funding}

This project has received funding from the European Research Council (ERC) under the ERC OxyPlanets projects (grant agreement No. 101053033). This project has received funding from the European Research Council (ERC) under the European Union's Horizon 2020 research and innovation programme (grant agreement No. 679030/WHIPLASH).

\section*{Acknowledgements}

A. Y. Jaziri acknowledges support from BELSPO BRAIN (B2/212/PI/PORTAL).

\section*{Author contributions statement}

A.Y. J. conceived of the study and performed the analysis. B. C performed analytical calculations to derive the critical value of Koxy. A.Y. J., N. C. and B. C. contributed to scientific discussion, interpretation of the results
and wrote the manuscript.

\section*{Competing interests}
The authors declare no competing interests.\\

\section*{Data availability}
Data associated to the GCM simulations have been upload to Zenodo with the following DOI: 10.5281/zenodo.7096149 and 10.5281/zenodo.14787904.\\

\section*{Code availability}
The code version associated to the GCM simulations has been upload to Zenodo with the following DOI: 10.5281/zenodo.7077747.\\
\noindent\textbf{Correspondence and requests for materials} should be addressed to Adam Y. Jaziri.

\clearpage

\section*{Appendix A: Transmission spectra with JWST NIRSpec Prism}
\label{an: spectra_prism}

\begin{figure}[ht]
\centering
\includegraphics[width=\linewidth]{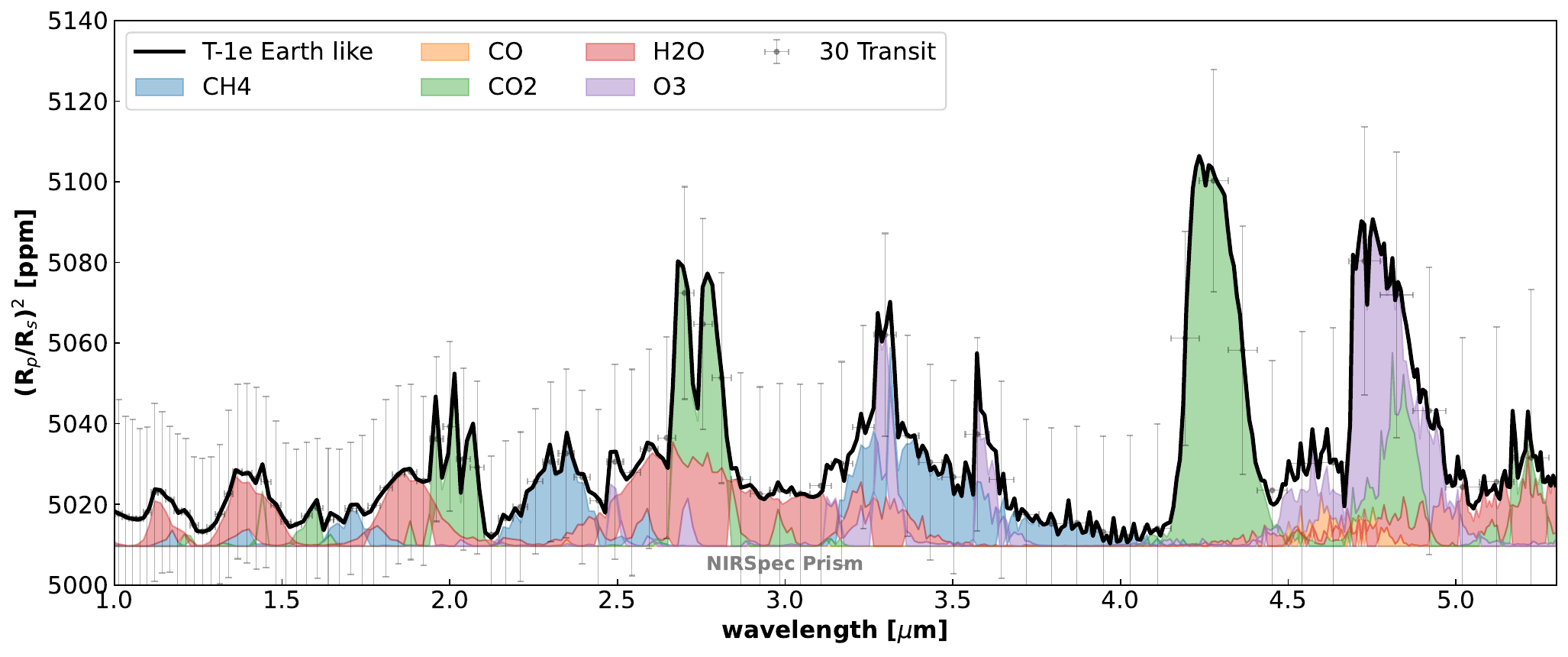}
\caption{TRAPPIST-1e transmission spectra for the Earth like simulation. Simulated noise of JWST observations for 30 transits with NIRSpec Prism is represented with the error bars at a resolution of 50. Molecules contributions are represented with shaded colors. Main O$_3$ features is around 4.6$\mu$m.}
\label{fig:spectra_bis}
\end{figure}

\clearpage

\section*{Appendix B: OH reaction rates}
\label{an: oh_rates}

\begin{figure}[ht]
\centering
\includegraphics[width=\linewidth]{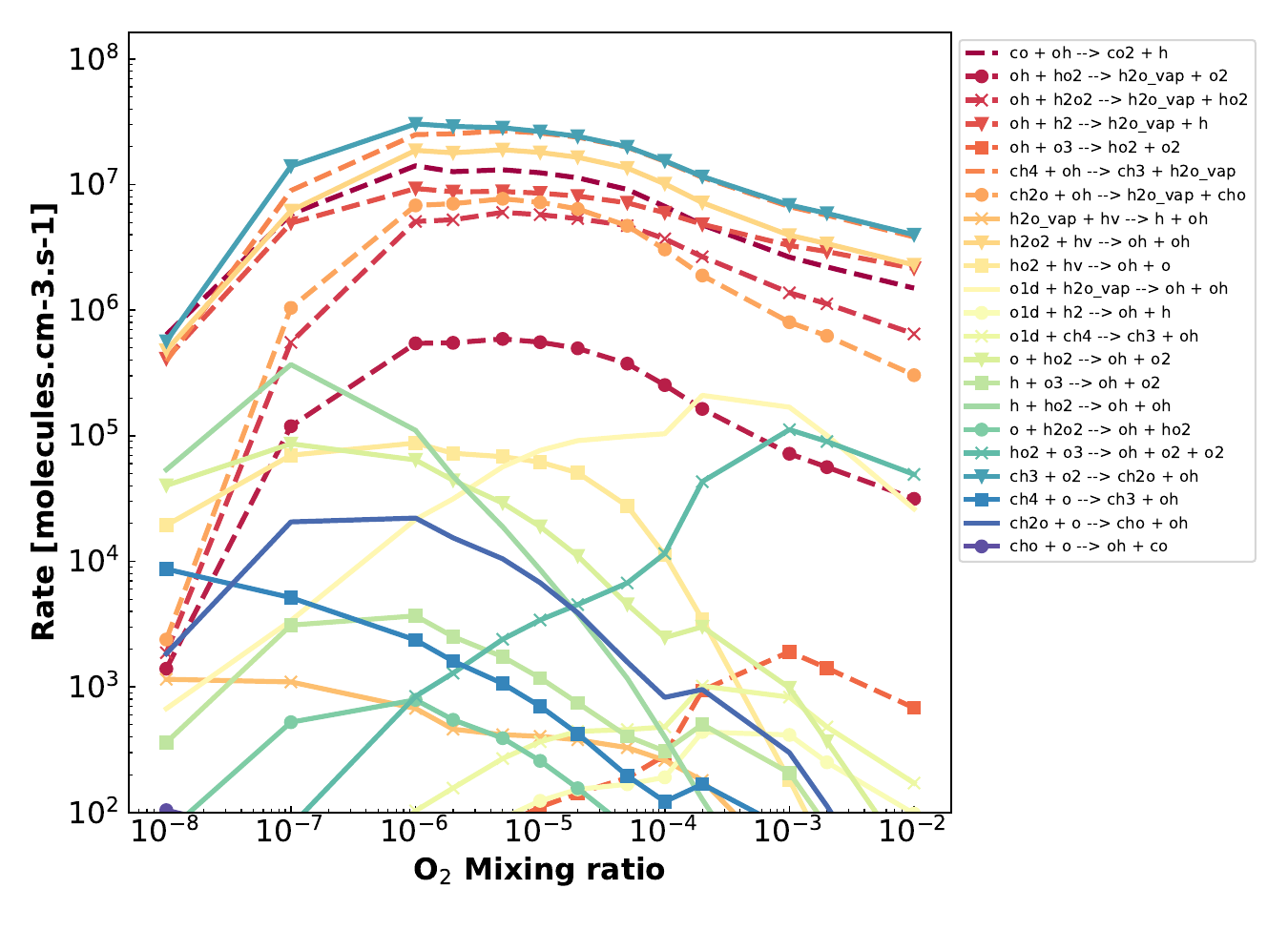}
\caption{OH reaction rates in the first layer of the 1D atmospheric model as a function of surface O$_2$. Only reactions with rates above 10$^{2}$ molecules·cm$^{-3}$·s$^{-1}$ are displayed. Solid lines represent OH production reactions, while dashed lines represent OH consumption reactions. The highest rates correspond to pathway (B1) (see Fig. \ref{fig:met_oxi}), highlighting methane oxidation as the dominant process. The secondary OH production rate originates from H$_2$O$_2$ photolysis, whereas H$_2$O photolysis is four orders of magnitude weaker.}
\label{fig:oh_rates}
\end{figure}

\clearpage

\section*{Appendix C: Evolution of odd oxygen}
\label{an: odd_oxygen}

\begin{figure}[ht]
\centering
\includegraphics[width=\linewidth]{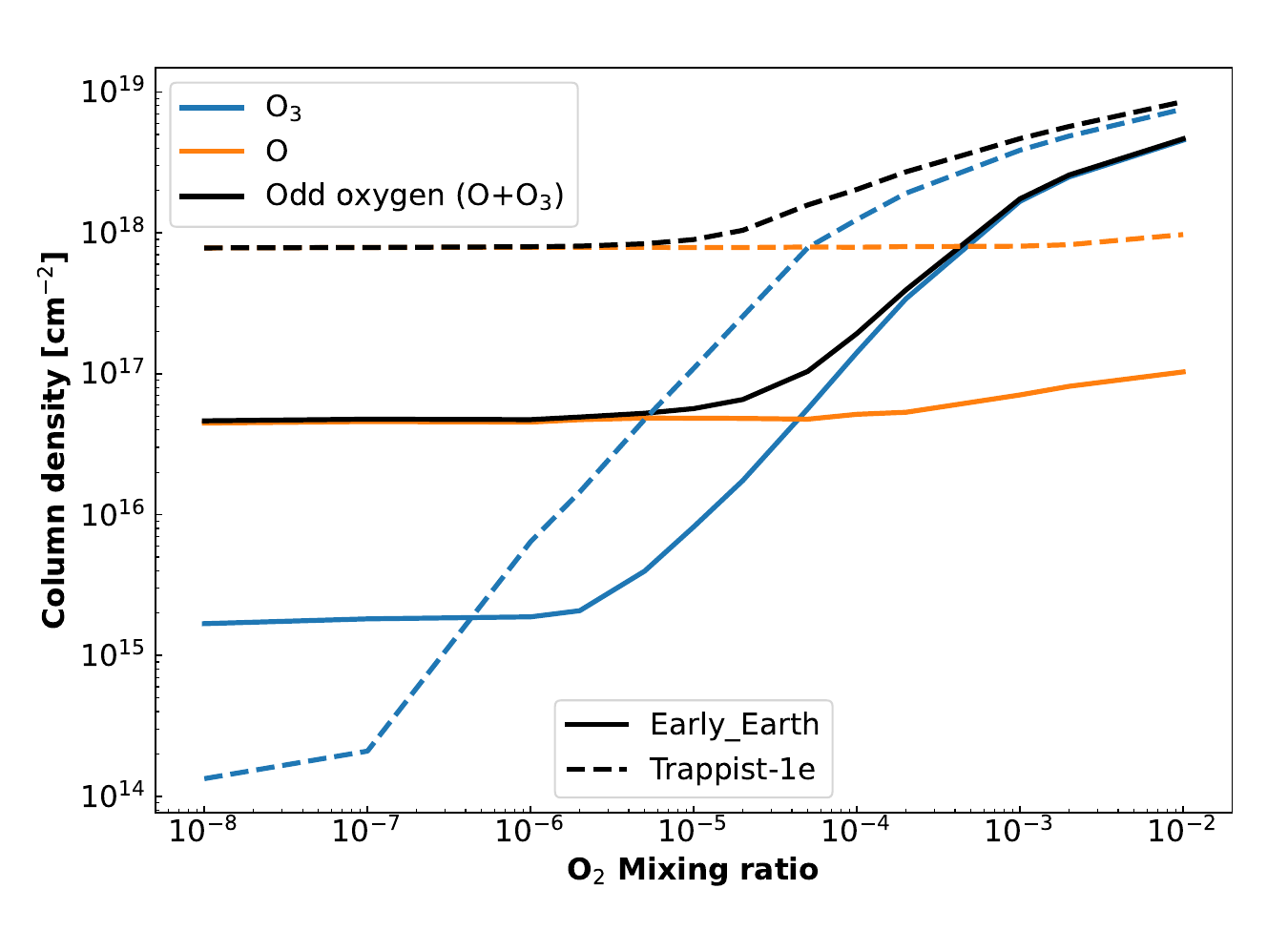}
\caption{Evolution of the total density column of odd oxygen (O + O$_3$) as a function of surface O$_2$. Results of early Earth model on Earth (solid) and on TRAPPIST-1e (dash).}
\label{fig:odd_oxygen}
\end{figure}

\end{document}